

\def\singlespace{\normalbaselines}
\def\oneandahalfspace{\baselineskip=1.15\normalbaselineskip plus 1pt
\lineskip=2pt\lineskiplimit=1pt}

\def\np{\vfill\eject}

\def\nofirstpagenoten{\nopagenumbers\footline={\ifnum\pageno>1\tenrm
\hss\folio\hss\fi}}
\def\nofirstpagenotwelve{\nopagenumbers\footline={\ifnum\pageno>1\twelverm
\hss\folio\hss\fi}}
\def\leaderfill{\leaders\hbox to 1em{\hss.\hss}\hfill}
\def\ft#1#2{{\textstyle{{#1}\over{#2}}}}
\def\frac#1/#2{\leavevmode\kern.1em
\raise.5ex\hbox{\the\scriptfont0 #1}\kern-.1em/\kern-.15em
\lower.25ex\hbox{\the\scriptfont0 #2}}
\def\sfrac#1/#2{\leavevmode\kern.1em
\raise.5ex\hbox{\the\scriptscriptfont0 #1}\kern-.1em/\kern-.15em
\lower.25ex\hbox{\the\scriptscriptfont0 #2}}


\parindent=20pt
\def\narrow{\advance\leftskip by 40pt \advance\rightskip by 40pt}

\def\AB{\bigskip
        \centerline{\bf ABSTRACT}\medskip\narrow}
\def\nonarrower{\advance\leftskip by -40pt\advance\rightskip by -40pt}
\def\AE{\bigskip\nonarrower}

\def\boxit#1{\vbox{\hrule\hbox{\vrule\kern3pt
        \vbox{\kern3pt#1\kern3pt}\kern3pt\vrule}\hrule}}

\def\gtorder{\mathrel{\raise.3ex\hbox{$>$}\mkern-14mu
             \lower0.6ex\hbox{$\sim$}}}
\def\ltorder{\mathrel{\raise.3ex\hbox{$<$}|mkern-14mu
             \lower0.6ex\hbox{\sim$}}}
\def\dalemb#1#2{{\vbox{\hrule height .#2pt
        \hbox{\vrule width.#2pt height#1pt \kern#1pt
                \vrule width.#2pt}
        \hrule height.#2pt}}}

\font\fourteentt=cmtt10 scaled \magstep2
\font\fourteenbf=cmbx12 scaled \magstep1
\font\fourteenrm=cmr12 scaled \magstep1
\font\fourteeni=cmmi12 scaled \magstep1
\font\fourteenss=cmss12 scaled \magstep1
\font\fourteensy=cmsy10 scaled \magstep2
\font\fourteensl=cmsl12 scaled \magstep1
\font\fourteenex=cmex10 scaled \magstep2
\font\fourteenit=cmti12 scaled \magstep1
\font\twelvett=cmtt10 scaled \magstep1 \font\twelvebf=cmbx12
\font\twelverm=cmr12 \font\twelvei=cmmi12
\font\twelvess=cmss12 \font\twelvesy=cmsy10 scaled \magstep1
\font\twelvesl=cmsl12 \font\twelveex=cmex10 scaled \magstep1
\font\twelveit=cmti12
\font\tenss=cmss10
 
 \font\ninebf=cmbx7 scaled \magstep1
\font\ninerm=cmr7 scaled \magstep1 \font\ninei=cmmi7 scaled \magstep1
\font\ninesy=cmsy7 scaled \magstep1 
\font\eightrm=cmr7 scaled 1140 
 
\font\sevenbf=cmbx7 \font\sevenrm=cmr7 \font\seveni=cmmi7
\font\sevensy=cmsy7 

\catcode`@=11
\newskip\ttglue
\newfam\ssfam

\def\fourteenpoint{\def\rm{\fam0\fourteenrm}
\textfont0=\fourteenrm \scriptfont0=\tenrm \scriptscriptfont0=\sevenrm
\textfont1=\fourteeni \scriptfont1=\teni \scriptscriptfont1=\seveni
\textfont2=\fourteensy \scriptfont2=\tensy \scriptscriptfont2=\sevensy
\textfont3=\fourteenex \scriptfont3=\fourteenex \scriptscriptfont3=\fourteenex
\def\it{\fam\itfam\fourteenit} \textfont\itfam=\fourteenit
\def\sl{\fam\slfam\fourteensl} \textfont\slfam=\fourteensl
\def\bf{\fam\bffam\fourteenbf} \textfont\bffam=\fourteenbf
\scriptfont\bffam=\tenbf \scriptscriptfont\bffam=\sevenbf
\def\tt{\fam\ttfam\fourteentt} \textfont\ttfam=\fourteentt
\def\ss{\fam\ssfam\fourteenss} \textfont\ssfam=\fourteenss
\tt \ttglue=.5em plus .25em minus .15em
\normalbaselineskip=16pt
\abovedisplayskip=16pt plus 4pt minus 12pt
\belowdisplayskip=16pt plus 4pt minus 12pt
\abovedisplayshortskip=0pt plus 4pt
\belowdisplayshortskip=9pt plus 4pt minus 6pt
\parskip=5pt plus 1.5pt
\setbox\strutbox=\hbox{\vrule height12pt depth5pt width0pt}
\let\sc=\tenrm
\let\big=\fourteenbig \normalbaselines\rm}
\def\fourteenbig#1{{\hbox{$\left#1\vbox to12pt{}\right.\n@space$}}}

\def\twelvepoint{\def\rm{\fam0\twelverm}
\textfont0=\twelverm \scriptfont0=\ninerm \scriptscriptfont0=\sevenrm
\textfont1=\twelvei \scriptfont1=\ninei \scriptscriptfont1=\seveni
\textfont2=\twelvesy \scriptfont2=\ninesy \scriptscriptfont2=\sevensy
\textfont3=\twelveex \scriptfont3=\twelveex \scriptscriptfont3=\twelveex
\def\it{\fam\itfam\twelveit} \textfont\itfam=\twelveit
\def\sl{\fam\slfam\twelvesl} \textfont\slfam=\twelvesl
\def\bf{\fam\bffam\twelvebf} \textfont\bffam=\twelvebf
\scriptfont\bffam=\ninebf \scriptscriptfont\bffam=\sevenbf
\def\tt{\fam\ttfam\twelvett} \textfont\ttfam=\twelvett
\def\ss{\fam\ssfam\twelvess} \textfont\ssfam=\twelvess
\tt \ttglue=.5em plus .25em minus .15em
\normalbaselineskip=14pt
\abovedisplayskip=14pt plus 3pt minus 10pt
\belowdisplayskip=14pt plus 3pt minus 10pt
\abovedisplayshortskip=0pt plus 3pt
\belowdisplayshortskip=8pt plus 3pt minus 5pt
\parskip=3pt plus 1.5pt
\setbox\strutbox=\hbox{\vrule height10pt depth4pt width0pt}
\let\sc=\ninerm
\let\big=\twelvebig \normalbaselines\rm}
\def\twelvebig#1{{\hbox{$\left#1\vbox to10pt{}\right.\n@space$}}}

\def\tenpoint{\def\rm{\fam0\tenrm}
\textfont0=\tenrm \scriptfont0=\sevenrm \scriptscriptfont0=\fiverm
\textfont1=\teni \scriptfont1=\seveni \scriptscriptfont1=\fivei
\textfont2=\tensy \scriptfont2=\sevensy \scriptscriptfont2=\fivesy
\textfont3=\tenex \scriptfont3=\tenex \scriptscriptfont3=\tenex
\def\it{\fam\itfam\tenit} \textfont\itfam=\tenit
\def\sl{\fam\slfam\tensl} \textfont\slfam=\tensl
\def\bf{\fam\bffam\tenbf} \textfont\bffam=\tenbf
\scriptfont\bffam=\sevenbf \scriptscriptfont\bffam=\fivebf
\def\tt{\fam\ttfam\tentt} \textfont\ttfam=\tentt
\def\ss{\fam\ssfam\tenss} \textfont\ssfam=\tenss
\tt \ttglue=.5em plus .25em minus .15em
\normalbaselineskip=12pt
\abovedisplayskip=12pt plus 3pt minus 9pt
\belowdisplayskip=12pt plus 3pt minus 9pt
\abovedisplayshortskip=0pt plus 3pt
\belowdisplayshortskip=7pt plus 3pt minus 4pt
\parskip=0.0pt plus 1.0pt
\setbox\strutbox=\hbox{\vrule height8.5pt depth3.5pt width0pt}
\let\sc=\eightrm
\let\big=\tenbig \normalbaselines\rm}
\def\tenbig#1{{\hbox{$\left#1\vbox to8.5pt{}\right.\n@space$}}}
\let\rawfootnote=\footnote \def\footnote#1#2{{\rm\parskip=0pt\rawfootnote{#1}
{#2\hfill\vrule height 0pt depth 6pt width 0pt}}}

\def\tenfoot{\tenpoint\hskip-\parindent\hskip-.1cm}

\overfullrule=0pt
\twelvepoint
\def\sbullet{\raise.2em\hbox{$\scriptscriptstyle\bullet$}}
\nofirstpagenotwelve
\hsize=16.5 truecm
\baselineskip 15pt

\def\ft#1#2{{\textstyle{{#1}\over{#2}}}}

\def\a{{\alpha_0^{\phantom \Phi}}}

\def\ve{\varepsilon}

\def\del{\partial}

\oneandahalfspace
\rightline{CTP TAMU--99/91}
\rightline{December 1991}

\vskip 2truecm
\centerline{\bf $N=2$ Superstrings with $(1,2m)$ Spacetime Signature}
\vskip 1.5truecm
\centerline{H. Lu,$^\star$ C.N. Pope,\footnote{$^
\star$}{\tenfoot Supported in part by the
U.S. Department of Energy, under
grant DE-FG05-91ER40633.} X.J. Wang and K.W.
Xu.\footnote{$^\$$}{\tenfoot Supported by a World Laboratory
Scholarship.}}
\vskip 1.5truecm
\centerline{\it Center
for Theoretical Physics,
Texas A\&M University,}
\centerline{\it College Station, TX 77843--4242, USA.}

\vskip 1.5truecm
\AB\singlespace

     We show that the $N=2$ superstring in $d=2D\ge6$ real dimensions, with
criticality achieved by including background charges in the two real time
directions, exhibits a ``coordinate-freezing'' phenomenon, whereby the
momentum in one of the two time directions is constrained to take a specific
value for each physical state.  This effectively removes this time direction
as a physical coordinate, leaving the theory with $(1,d-2)$ real spacetime
signature. Norm calculations for low-lying physical states suggest that the
theory is ghost free.

\AE\oneandahalfspace

\np
\noindent
{\bf }
\bigskip

      The $N=2$ superstring is usually considered to be unattractive physically
for a number of reasons.  One of these is that although its critical
dimension is $4$, the $N=2$ supersymmetry implies the existence of a complex
structure [1], which means that the spacetime signature cannot be Minkowskian;
instead it must be $(0,4)$, $(2,2)$ or $(4,0)$.  The critical dimension $4$
reflects the fact that the critical central charge for the $N=2$
super-Virasoro algebra is $c=6$. This can be realised by two $N=2$
superfields without background charges, thus giving the usual $N=2$
superstring in two complex dimensions [2,3].  In this paper, we study more
general realisations with arbitrary numbers of $N=2$ superfields and with
background charges which are fixed so that the theories are critical.
This gives rise to critical $N=2$ superstrings in arbitrary numbers of
complex dimensions. We show that the existence of background charges will
freeze one real coordinate, {\it i.e.} the momentum component in that
direction is constrained by the physical-state conditions to take a specific
value for each state.  If we start from a theory with one complex time and
choose the background charge to lie in that direction, then the result of
the coordinate freezing is to give a theory that effectively has only one
real time direction.  Thus although $N=2$ superstrings suffer from a number
of phenomenological defects, the spacetime signature of the critical
theories in more than four real dimensions can, at least, effectively be
Minkowskian, {\it i.e.} $(1,2m)$ with $m \ge 2$.  Therefore we shall assume
in the rest of this paper that the background charge is taken to lie in the
complex time direction. We also show in this paper that the norms of
low-lying physical states are non-negative, suggesting that the theories are
ghost free.

         In component form, the currents for the $N=2$ super-Virasoro algebra
are given by
$$
\eqalign{
J&=-\psi^\mu\bar\psi_\mu - \a \del\phi_0 +\a \del\bar\phi_0 \ ,\cr
G^+&=\sqrt2(\del \bar\phi^\mu\psi_\mu- \a\del\psi_0)\ ,\cr
G^-&=\sqrt2(\del\phi^\mu\bar\psi_\mu - \a\del\bar\psi_0)\ , \cr
T&=\ft12\psi^\mu\del\bar\psi_\mu - \ft12\del\psi^\mu\bar\psi_\mu -
\del\phi^\mu\del\bar\phi_\mu + \ft12\a\del^2\phi_0 +
\ft12\a\del^2\bar\phi_0\ , \cr}\eqno(1)
$$
where we have chosen the background charge to lie in the $\mu=0$ direction.
The index $\mu$ runs over $0,1,\ldots,D-1$. The central charge of this
realisation is
$$
c=3(D-2\alpha^2_0)\ .\eqno(2)
$$
The anomaly-freedom condition $c=6$ therefore requires that
$$
\alpha^2_0=\ft12(D-2)\ .\eqno(3)
$$

  	A physical state $|p\big\rangle$ satisfies the physical-state
conditions
$$
\eqalign{
L_m\big|p\big\rangle&=0=J_m\big|p\big\rangle \qquad m\ge 0\ ,\cr
G^+_r\big|p\big\rangle&=0=G^-_r\big|p\big\rangle \qquad r>0\ .\cr}
\eqno(4)
$$
(Since the intercepts of $J_0$ and $L_0$ are zero for the $N=2$
super-conformal algebra, we include these in the physical-state conditions
(4).) Physical states can be constructed by acting on an $SL(2,C)$-invariant
vacuum $\big|0\big\rangle$ with ground-state operators $P(z)$, {\it i.e.}\
$\big|p\big\rangle\equiv P(0)\big|0\big\rangle$.  The ground-state operators
take the form
$$
P(z)=R(z) e^{\beta\cdot\bar\phi+\bar\beta\cdot\phi}\ .\eqno(5)
$$
(Normal ordering is understood.)  The operators $R(z)$ can be classified by
their eigenvalues $\ell$ and $n$ under $J_0$ and $L_0$ respectively.  The
eigenvalue $\ell$ measures the fermion charge of the operator $R(z)$; each
$\psi_\mu$ in a monomial in $R(z)$ contributes $+1$, each ${\bar\psi}_\mu$
contributes $-1$, and $\del\phi_\mu$ and $\del\bar\phi_\mu$ contribute 0.
The eigenvalue $n$ measures the conformal dimension of the operator
$R(z)$, {\it i.e.}\ the level number.

    At level $n=0$, $R$ is just the identity operator, with $\ell=0$, and
$P(z)$ is the ``tachyon" ground-state operator. At level $n=\ft12$, $R$
can be $\bar\xi_\mu\psi^\mu$, with $\ell=+1$; or $\xi_\mu\bar\psi^\mu$, with
$\ell=-1$. At level $n=1$, $\ell$ can be $-2,\ 0,\ +2$. In general, at level
$n$, $\ell$ takes the values
$$
\ell =-2n,\ -2n+2,\ \ldots,\ 2n-2,\ 2n\ .\eqno(6)
$$

      It is convenient to work with $2D$ real coordinates $\varphi^\mu$
and $\widetilde\varphi^\mu$,  rather than the $D$ complex
coordinates $\phi^\mu$, where $\mu=0,1,\ldots,D-1$. Thus we define
$$
\eqalign{
\phi^\mu&={1\over{\sqrt 2}}\big(\varphi^\mu+i\widetilde\varphi^\mu\big)\ ,\cr
\bar\phi^\mu&={1\over{\sqrt 2}}\big(\varphi^\mu-i\widetilde\varphi^\mu\big)
\ .\cr}\eqno(7)
$$
Correspondingly, we can introduce $2D$ parameters $k_\mu$ and $\tilde
k_\mu$ which are defined by
$$
\eqalign{
\beta_\mu&={i\over{\sqrt 2}}\big(k_\mu+i \tilde k_\mu\big)\ ,\cr
\bar\beta_\mu&={i\over{\sqrt 2}}\big(k_\mu-i \tilde k_\mu\big)\ .\cr}
\eqno(8)
$$
It follows that
$$
e^{\bar\beta\cdot\phi+\beta\cdot\bar\phi}
=e^{ik\cdot\varphi+i{\tilde k}\cdot{\widetilde\varphi}}
\ .\eqno(9)
$$
Thus $k_\mu$ and $\tilde k_\mu$ are the momenta conjugate to $\varphi^\mu$
and $\widetilde\varphi^\mu$ respectively.

    For physical states with level number $n$ and fermion
charge $\ell$, the $J_0$ and $L_0$ constraints in (4) give
$$
\eqalignno{
J_0:\qquad  0&=\ell+\a (\beta_0-\bar\beta_0)\cr
&=\ell-\sqrt2 \a \tilde k_0\ ,&(10a)\cr\cr
L_0:\qquad 0&=n -\beta^\mu \bar\beta_\mu +\ft12 \a (\beta_0+\bar\beta_0)\cr
&=n+\ft12 k^\mu k_\mu +\ft12 \tilde k^\mu \tilde k_\mu + \ft{i}{\sqrt2}\a k_0
\ .&(10b)\cr}
$$
In the usual discussion of the $N=2$ superstring, for which $D=2$ and hence
from (2) the background charge $\a$ is zero, equation (10$a$) implies
that for all physical states the fermionic charge of $R(z)$ must be zero.
This implies that in this case there are no physical states occurring at
levels with $n$ a half-integer.  (In fact, as discussed in [2,3], all the
higher-level physical states are longitudinal, and hence have no physical
degrees of freedom.)

    We are interested in the case where $D$ is greater than 2, which
implies, from (2), that $\a$ is real. It follows from equation (10$a$) that the
momentum component $\tilde k_0$ is ``frozen'' to the value
$$
\tilde k_0= {\ell\over {\sqrt2 \ \a}}\ .\eqno(11)
$$
(A similar momentum-freezing phenomenon was found for the bosonic $W_3$
string in [4].  In fact it seems to be the case that momentum freezing
occurs for any string theory based on an extended conformal
algebra with additional bosonic currents, when background charges are
present.)

        The hermiticity conditions for $L_0$ and $J_0$ imply that
$$
\eqalign{
k^{*}_i&=k_i \qquad i=1,2,\ldots,D-1 \ , \cr
\tilde k^{*}_\mu&={\tilde k}_\mu \qquad \mu=0,1,\ldots,D-1 \ , \cr} \eqno(12)
$$
where ``$*$" denotes complex conjugation;
whilst $k_0$ must satisfy a condition that is modified by
the background charge:
$$
k^*_0=k_0-i\sqrt2 \a \ .\eqno(13)
$$
For this, it is convenient to introduce a shifted momentum $\hat k_0$ that is
real:
$$
\hat k_0=k_0 -{i\over \sqrt2} \a \ ,\eqno(14)
$$
It follows that the complex-conjugation relations for $\beta_\mu$ and
$\bar\beta_\mu$ are
$$
\eqalignno{
\beta_i^*&=-\bar\beta_i\ ;\qquad \qquad{\bar\beta}_i^*=-\beta_i\ , &(15a)\cr
\beta_0^*&=-\bar\beta_0-\a\ ;\qquad {\bar\beta}_0^*=-\beta_0-\a\ ,
&(15b)\cr}
$$
where the index $i=1,2,\cdots,D-1$.

     For the theory to be unitary, the norms of the physical states should all
be non-negative. For the case of the $N=2$ superstring in $D=2$ complex
dimensions, for which background charges vanish, the no-ghost theorem has
been discussed in [3,5]. For the case of interest in this paper, $D>2$, we
shall show explicitly that at the levels $n=\ft12$ and $n=1$, the norms for the
physical states are non-negative, too.  Usually, it seems to be the case
that the absence of negative-norm states at low-lying levels provides a
reliable
indication of the ghost freedom of the theory.

     At level $n=\ft12$, there are vector states which have fermion charge
$\ell$ equal to $+1$ or $-1$.
Without loss of generality, we shall focus on the $\ell=+1$ states, for
which the ground-state operator is given by (5) with
$$
R(z)=\bar\xi_\mu\psi^\mu(z)\ .\eqno(16)
$$
In addition to the $J_0$ and $L_0$ constraints, as given in (10$a$,$b$),
there is, in this case, one other nontrivial constraint implied by the
physical-state conditions (4), namely $G^-_{1/2}\big|p\big\rangle=0$. This
implies
$$
{\bar\xi}_\mu\beta^\mu-{\bar\xi}_0\a=0\ .\eqno(17)
$$
{}From this, we can solve for $\bar\xi_0$ in terms of $\bar\xi_i$:
$$
\bar\xi_0={{\bar\xi_i\beta_i}\over{\beta_0+\a}}\ .\eqno(18)
$$
The norm $\cal N$ of these states is
$$
{\cal N}=\bar\xi_\mu^*\bar\xi^\mu=-\bar\xi_0^*\bar\xi_0+
\bar\xi_i^*\bar\xi_i\ .\eqno(19)
$$
Substituting (18) into (19), we can write $\cal N$ as
$$\eqalign{
{\cal N}&=\bar\xi_i^*\Big(\delta_{ij}-V_i^*V_j\Big)\bar\xi_j\cr
&\equiv  \bar\xi_i^* N_{ij}\bar\xi_j\ ,\cr}\eqno(20)
$$
where $i=1,\ 2,\ \ldots,\ D-1$ and
$$
V_i={\beta_i\over{\beta_0+\a}}\ .\eqno(21)
$$
It is easy to see that there are $D-2$ transverse eigenvectors of $N_{ij}$
with eigenvalue $+1$, whilst the eigenvector parallel to $V_i$ has
eigenvalue $1-V_i^*V_i$, which is zero by virtue of the $J_0$ and $L_0$
constraints given in (10$a$) and (10$b$).  It follows that of the $D-1$
independent complex polarisation states, one has zero norm, whilst the
remaining $D-2$ states have positive norm. A completely analogous discussion
can be given for the $\ell=-1$ vector states at level $n=\ft12$, leading to
the same result for the numbers of postive-norm and zero-norm states.

     At level $n=1$ we know from equation (6) that the fermion charge $\ell$
can be $-2,\ 0,\ {\rm or}\ +2$. The $\ell=-2$ case essentially gives the
same result for the norms as the $\ell=+2$ case; we shall therefore just
consider the $\ell=+2$ and $\ell=0$ cases. For $\ell=+2$, the ground-state
operator $R(z)$ in equation (5) takes the form
$$
R(z)=\bar\xi_{\mu\nu}\psi^\mu\psi^\nu\ .\eqno(22)
$$
In addition to the $J_0$ and $L_0$ constraints,  there is again just one
other nontrivial constraint, namely $G^-_{1/2}\big|p\big\rangle=0$. This
implies that the antisymmetric polarisation tensor $\bar\xi_{\mu\nu}$ must
satisfy the following transversality condition:
$$
\bar\beta_0^*\bar\xi_{0\nu}+\beta_i\bar\xi_{i\nu}=0\ .\eqno(23)
$$
Thus we can solve for $\bar\xi_{0i}$ in terms of $\bar\xi_{ij}$, and
substitute this into the norm $\bar\xi_{\mu\nu}^*{\bar\xi}^{\mu\nu}$ to
show that it is positive semi-definite: Of the $\ft12(D-1)(D-2)$ independent
complex components of the polarisation tensor $\bar\xi_{\mu\nu}$, we find
that $D-2$ give rise to zero-norm states, whilst the the remaining
$\ft12(D-2) (D-3)$ give states of positive norm.

     For $\ell=0$ states at level $n=1$, $R(z)$ of equation (5) takes the form
$$
R(z)=\ve_{\mu\nu}\psi^\mu\bar\psi^\nu
+\bar\xi_\mu\del\phi^\mu+\xi_\mu\del\bar\phi^\mu\ .\eqno(24)
$$
In this case, in addition to the $J_0$ and $L_0$ constraints, we have three
other independent nontrivial constraints, coming from $J_1$, $G^+_{1/2}$ and
$G^-_{1/2}$. They give, respectively,
$$
\eqalignno{
\ve^\mu{}_\mu&=\a(\xi_0-\bar\xi_0)\ ,&(25a)\cr
\bar\xi^\mu&=\beta_\nu^*\ \ve^{\mu\nu}\ ,&(25b)\cr
\xi^\mu&=-{\bar\beta}_\nu^*\ \ve^{\nu\mu}\ .&(25c)\cr}
$$

      The norm $\cal N$ for these states is given by
$$
{\cal N}=\ve^*_{\mu\nu}\ve^{\mu\nu}+{\bar\xi}^*_\mu{\bar\xi}^\mu+
{\xi}^*_\mu{\xi}^\mu\ .\eqno(26)
$$
{}From (25$b$) and (25$c$) we may eliminate $\bar\xi^\mu$ and $\xi^\mu$ in
(26), and express the norm purely in terms of $\ve_{\mu\nu}$, subject to the
constraint implied by (25$a$). It is convenient to decompose $\ve_{\mu\nu}$
into irreducible transverse and traceless parts, in the following manner:
$$
\ve_{\mu\nu}=\ve^{TT}_{\mu\nu}+{\bar\ve}^T_\mu\beta_\nu
+\bar\beta_\mu\ve^T_\nu +\lambda\bar\beta_\mu\beta_\nu
+\kappa\eta_{\mu\nu}\ ,\eqno(27)
$$
where
$$\eqalign{
\ve^{TT\mu}_\mu&=0\ ,\quad \ve^{TT}_{\mu\nu}\beta^{\nu *}=0\ ,
\quad {\bar\beta}^{\mu *}\ve^{TT}_{\mu\nu}=0\ ;\cr
\ve^T_\mu\beta^{\mu *}&=0\ ,\quad \bar\beta^{\mu *}\bar\ve^T_\mu=0\ .\cr}
\eqno(28)
$$
Substituting (27) into (25$a$-$c$), we find that $\kappa $ and $\lambda$ are
related by
$$
\ft13(c-6)\kappa+(\lambda+2\kappa)\Big(1-\ft12\a(\beta_0+\bar\beta_0)\Big)
=0\ ,\eqno(29)
$$
where we have made use of equations (2), (10$a$,$b$) and (15$a$,$b$). Thus
$\lambda$ and $\kappa$ are not independent, and together represent just one
complex state. Now we find that the norm $\cal N$ is given by
$$
{\cal N}={\ve^{TT*}_{\mu\nu}}\ve^{TT\mu\nu}+\ft13 (6-c)\kappa^*\kappa
-(\lambda+2\kappa)^*(\lambda+2\kappa)
\ .\eqno(30)
$$
The criticality condition $c=6$ implies, from equation (29), that $\lambda+
2\kappa=0$. Consequently, the last two terms in (30) vanish, showing that
the corresponding state has zero norm. The absence of $\ve^T_\mu$ and
$\bar\ve^T_\mu$ in (30) implies that these components of the decomposition
(27) of $\ve_{\mu\nu}$ are associated with zero-norm states, numbering $2D-2$
in all.

      Solving for $\ve^{TT}_{00}$, $\ve^{TT}_{0i}$ and $\ve^{TT}_{i0}$,
using (28), the remaining terms in (30) give
$$
{\cal N}=\ve^{TT*}_{ij}\ve^{TT}_{ij}
-\ve^{TT*}_{ij}\ve^{TT}_{ik}{\bar V}^*_j{\bar V}_k
-\ve^{TT*}_{ji}\ve^{TT}_{ki}V^*_jV_k
+\ve^{TT*}_{ij}\ve^{TT}_{kl}V^*_i{\bar V}^*_jV_k{\bar V}_l\ ,\eqno(31)
$$
where $i,\ j,\ k,\ l=1,\ 2,\ \ldots,\ D-1$, $V_i$ is given in (21)
and $\bar V_i$ is given by an analogous expression with $\beta_\mu$ replaced
by $\bar\beta_\mu$.  It is straightforward to see that of the $(D-1)^2$
complex states, the $(D-2)^2$ transverse states have positive norm,
whilst the remaining $2D-3$ states have zero norm.

      With these results, we have demonstrated the absence of negative-norm
states at levels $n=\ft12$ and $n=1$.  Although we have not considered a
general no-ghost theorem here, we would expect that if negative-norm states
were to occur, they would have arisen already at the levels that we have
examined. The theories we are considering here differ qualitatively from the
usual $D=2$ superstring, for which the no-ghost theorem has been discussed
[3,5], since the extra spatial dimensions allow the existence of
positive-norm as well as zero-norm physical states at levels $n>0$. As usual
in string theory, the occurrence of the zero-norm states indicates that the
physical states at level $n>0$ describe gauge fields. Note also that when
$D>2$ not only do positive-norm excited states occur, but also the level number
$n$ can take half-integer as well as integer values.  This contrasts with
the $D=2$ case where, since $\a=0$, equations (6) and (10$a$) imply that $n$
can take only integer values.

     The most striking feature arising in the $D>2$ case is that the momentum
component $\tilde k_0$ in the direction of the real time coordinate $\tilde
\varphi^0$ is frozen to a specific value for each physical state, as given
by (11).  Because of this, the coordinate $\tilde\varphi^0$ does not
describe an observable spacetime dimension. (This is somewhat analogous to
Kaluza-Klein dimensional reduction, where one truncates to the zero-momentum
mode in the Fourier expansion of the coordinate dependence for a
compactified coordinate, thereby obtaining a theory in one less spacetime
dimension.) Thus we are effectively left with just one real time dimension,
parametrised by the coordinate $\varphi^0$.  Note that this
momentum-freezing phenomenon does not occur in the usual $N=2$ superstring
in $D=2$ complex dimensions, since then $\a$ is zero and so the momentum
component $\tilde k_0$ will not be constrained by (10$a$).

        In a system which is Poincar\'e invariant, mass is defined by the
Poincar\'e Casimir operator $-P^\mu P_\mu$. For $N=2$ superstring theories,
the spacetime has a complex structure, and the Lorentz subgroup $SO(2,2D-2)$
is replaced by $U(1,D-1)$. When $D>2$, the presence of the background charges
will break this symmetry further. Thus one cannot have a clear definition of
the ``mass" such as the Poincar\'e Casimir operator. In this case, from the
mass-shell condition (10$b$) there are two natural mass-type operators that
one might define [4]: ${\cal M}$ and ${\widetilde {\cal M}}$,
given by
$$
\eqalignno{
{\cal M}^2&\equiv -k_\mu^*k^\mu-{\tilde k}_i^*{\tilde k}_i\ ,&(32a)\cr
{\widetilde{\cal M}}^2&\equiv {\hat k}_0{\hat k}_0-k_i^*k_i
-{\tilde k}_i^*{\tilde k}_i\ .&(32b)\cr}
$$
It follows from (10$a$) and (10$b$) that the spectra for $\cal M$ and
$\widetilde{\cal M}$ are
$$
\eqalignno{
{\cal M}^2&=2n-{\ell^2\over{2\alpha^2_0}}\ ,&(33a)\cr
{\widetilde{\cal M}^2}&=2n-{\ell^2\over{2\alpha^2_0}}-\ft12\alpha^2_0
\ .&(33b)\cr}
$$
For now, we reserve judgement on which, if either, of these provides a
useful generalisation of the notion of mass.  Note that these formulae
would suggest that the theories would contain infinite numbers of
arbitrarily massive tachyonic states corresponding to sufficiently large
$\ell$ values, within the range given by (6), at given level $n$.  This
problem could be overcome by truncating the states of the theory to just the
$\ell=0$ sector. It is worth noting that this truncation would have the
consequences that only levels with $n$ equal to an integer would survive,
and also that the momentum $\tilde k_0$ of all surviving physical states
would be frozen to the value zero.

\bigskip\bigskip
\bigskip\bigskip
\singlespace
\centerline{\bf ACKNOWLEDGMENT}
\frenchspacing
\bigskip

       We are very grateful to HoSeong La for his meticulous reading of the
manuscript.

\bigskip\bigskip
\bigskip\bigskip
\singlespace
\centerline{\bf REFERENCES}
\frenchspacing
\bigskip

\item{[1]}L. Alvarez-Gaum\'e and D.Z. Freedman, Comm. Math. Phys. {\bf 80}
(1981) 443.

\item{[2]}M.B.\ Green, J.H.\ Schwarz and E.\ Witten, ``Superstring Theory,''
(CUP 1987).

\item{[3]}H. Ooguri and C. Vafa, Nucl. Phys. {\bf B361} (1991) 469.

\item{[4]}C.N.\ Pope, L.J.\ Romans E.\ Sezgin and K.S.\ Stelle, ``The $W_3$
String Spectrum,'' preprint CTP TAMU-68/91.

\item{[5]}J. Bie\'nkowska, ``The generalised no-ghost theorem for $N=2$ SUSY
critical strings,'' preprint, EFI 91-65.

\end